\documentclass[twocolumn]{aastex631}
\usepackage{multirow}

\graphicspath{{./}{figures/}}

\begin{document}

\title{ Multi-frequency VLBI Observations of the M\,84 Inner Jet/Counterjet }

\author{Xuezheng Wang}
\affiliation{Shanghai Astronomical Observatory, Chinese Academy of Sciences, 80 Nandan Road, Shanghai 200030, People's Republic of China; \href{jiangwu@shao.ac.cn}{jiangwu@shao.ac.cn}} 
\affiliation{ShanghaiTech University, 393 Middle Huaxia Road, Pudong, Shanghai, 201210, People's Republic of China}
\affiliation{University of Chinese Academy of Sciences, No.19(A) Yuquan Road, Shijingshan District, Beijing,100049, People's Republic of China}
\author[0000-0001-7369-3539]{Wu Jiang}
\affiliation{Shanghai Astronomical Observatory, Chinese Academy of Sciences, 80 Nandan Road, Shanghai 200030, People's Republic of China; \href{jiangwu@shao.ac.cn}{jiangwu@shao.ac.cn}}
\affiliation{Key Laboratory of Radio Astronomy, Chinese Academy of Sciences, 210008 Nanjing, People's Republic of China}

\author[0000-0003-3540-8746]{Zhiqiang Shen}
\affiliation{Shanghai Astronomical Observatory, Chinese Academy of Sciences, 80 Nandan Road, Shanghai 200030, People's Republic of China; \href{jiangwu@shao.ac.cn}{jiangwu@shao.ac.cn}}
\affiliation{Key Laboratory of Radio Astronomy, Chinese Academy of Sciences, 210008 Nanjing, People's Republic of China}
\author[0000-0002-1923-227X]{Lei Huang}
\affiliation{Shanghai Astronomical Observatory, Chinese Academy of Sciences, 80 Nandan Road, Shanghai 200030, People's Republic of China; \href{jiangwu@shao.ac.cn}{jiangwu@shao.ac.cn}}
\affiliation{Key Laboratory for Research in Galaxies and Cosmology, Shanghai Astronomical Observatory, Chinese Academy of Sciences, Shanghai, 200030, People's Republic of China}
\author[0000-0001-6906-772X]{Kazuhiro Hada}
\affiliation{Mizusawa VLBI Observatory, National Astronomical Observatory of Japan, 2-12 Hoshigaoka, Mizusawa, Oshu, Iwate 023-0861, Japan}
\affiliation{Department of Astronomical Science, The Graduate University for Advanced Studies (SOKENDAI), 2-21-1 Osawa, Mitaka, Tokyo 181-8588, Japan}

\author[0000-0001-6311-4345]{Yuzhu Cui}
\affiliation{Tsung-Dao Lee Institute Shanghai Jiao Tong University 520 Shengrong Road, Shanghai 201210, People's Republic of China}
\author[0000-0002-7692-7967]{Ru-Sen Lu}
\affiliation{Shanghai Astronomical Observatory, Chinese Academy of Sciences, 80 Nandan Road, Shanghai 200030, People's Republic of China; \href{jiangwu@shao.ac.cn}{jiangwu@shao.ac.cn}}
\affiliation{Key Laboratory of Radio Astronomy, Chinese Academy of Sciences, 210008 Nanjing, People's Republic of China}
\affiliation{Max-Planck-Institut f\"ur Radioastronomie, Bonn, Germany}


\begin{abstract}

\noindent Observational studies of inner-most regions of the edge-on jets in nearby active galactic nuclei (AGN) are crucial to understand their kinematics and morphology. For the inner jet of the nearby low-luminosity AGN in M\,84, we present new high-sensitivity observations with very long baseline interferometry since 2019, as well as archival Very Long Baseline Array observations in 2014.
We find that the compact core in M\,84 has an inverted-to-flat spectrum from 1.5 to 88\,GHz. Based on the turnover frequency of $4.2\pm0.2$\,GHz in the spectrum, we estimated a magnetic field strength of $1 \sim 10$\,mG and an electron number density of $\sim 10^5\, {\rm cm^{-3}}$ in the core region. 
Three inner jet components within $\sim$3\,mas from the core are identified and traced in the images at 22\,GHz, whose apparent speeds are 0.11\,c, 0.27\,c, and 0.32\,c, respectively. We calculate the viewing angle of ${58^{+17}_{-18}}^{\circ}$ for the inner jet based on the proper motion and the flux ratio of jet-to-counterjet. A propagating sinusoidal model with a wavelength of $\sim$3.4\,mas is used to fit the helical morphology of the jet extended to 20\,mas ($\sim 2.2 \times 10^4$ Schwarzschild Radii).

\end{abstract}

\keywords{\href{http://astrothesaurus.org/uat/1390}{ Relativistic jets (1390)}; \href{http://astrothesaurus.org/uat/2033}{Low-luminosity active galactic nuclei(2033)};        \href{http://astrothesaurus.org/uat/1769}{Very long baseline interferometry(1769)}}

\section{Introduction} \label{sec:intro}

\newcommand{\tabincell}[2]{\begin{tabular}{@{}#1@{}}#2\end{tabular}} 
\begin{deluxetable*}{ccccccc}
\tablenum{1}
\tablecaption{Image parameters \label{tab:image}}
\tablewidth{0pt}
\tablehead{
\colhead{Code} &  \colhead{Array} & \colhead{Date (yyyy/mm/dd)}& \colhead{Frequency (GHz)}  & \colhead{Beam (mas $\times$ mas, deg)}  & \colhead{Core flux (Jy)} & RMS (mJy/beam) 
}
\decimalcolnumbers
\startdata
bj094a  & VLBA & 2019/06/22& 22.0 &  1.37$\times$0.40, $-$13.8  & 0.086$\pm$0.011 & 0.11 \\
a19xw01a & EAVN &2019/12/20 & 22.2   & 1.44$\times$0.61, 17.8 &  0.101$\pm$0.011 & 0.11  \\
\multirow{2}{*}{bx014}& \multirow{2}{*}{VLBA}  &\multirow{2}{*}{2020/06/02}&  4.9 (b) & 3.46$\times$1.65, $-$5.1 &  0.125$\pm$0.012 & 0.04  \\
								&  &  & 23.6 (d) & 0.78$\times$0.34, $-$7.6 &  0.086$\pm$0.012 & 0.08 \\ 
\multirow{2}{*}{bj094b} & \multirow{2}{*}{VLBA} & \multirow{2}{*}{2021/03/31}& 44.1 (e) & 0.56$\times$0.23, $-$17.8 &  0.086$\pm$0.012 & 0.15\\
									& 		 & &  87.8 (f) & 0.27$\times$0.11, $-$17.1 	 & 0.083$\pm$0.016 & 0.70\\
\hline
\multirow{3}{*}{bh186d*} & \multirow{3}{*}{VLBA}  & \multirow{3}{*}{2014/03/26}&1.5 (a) & 12.9$\times$6.43, 16.6  & 0.061$\pm$0.009 & 0.08\\
								& 		&  &	5.0 & 3.86$\times$1.99, 18.3 	&	0.080$\pm$0.009  &  0.07 \\
								&		& &		15.5 (c) &  1.26$\times$0.65, 19.0 & 0.093$\pm$0.012 	& 0.11 \\
\multirow{3}{*}{bh186e*}& \multirow{3}{*}{VLBA} 	& \multirow{3}{*}{2014/05/08} & 1.5 & 10.4$\times$4.97, $-$1.3  &  0.067$\pm$0.008 	& 0.06 \\
								&  	&&  5.0 & 3.17$\times$1.61, $-$1.4   &  0.091$\pm$0.010 	& 0.06 \\
								& 		&  & 	15.5 &  1.07$\times$0.55, $-$4.2	& 0.080$\pm$0.011 &0.10\\
\enddata
\tablecomments{ Columns (1)$\sim$(5): VLBA/EAVN legacy experiment code (* means the data from public archive), VLBI array, observing date, frequency, and full width at half maximum of the synthesized beam under natural weighting, respectively. The alphabets in the brakets listed the frequency column correspond to the same labels in Figure \ref{fig:image}; (6) The flux of the core component; (7) The root mean square (RMS) noise of images. }
\end{deluxetable*}
M\,84 (NGC\,4374, 3C\,272.1) ($z$ = 0.00339, $D$ = 18.4\,Mpc) is an elliptical galaxy in the Virgo cluster with a low luminosity active galaxy nucleus (LLAGN). The central supermassive black hole, with a mass of $8.5\times10^8$$M_\odot$ measured by the gas kinematics \citep{Walsh}, or $~1.8\times10^9$$M_\odot$ estimated from the velocity dispersion \citep{Ly}, launches a weak Fanaroff-Riley type I radio jet. A two-side jet was observed at radio and UV wavebands \citep{Meyer}. A large viewing angle of $74^{+9}_{-18}$$^\circ$ was calculated based on the outer jet at hundreds of parsec scale with its jet-to-counterjet flux ratio and apparent speed. 
The inner jet at sub-parsec scale near the compact core has been studied in the radio band using the very long baseline interferometry (VLBI) technique. Very Long Baseline Array (VLBA) observation at 5\,GHz clearly showed two-sided extended structures \citep{Nagar}. A high angular resolution VLBI image at 43\,GHz exhibited a compact core with a slight extension to the north \citep{Ly}. \citet{Nakahara} adopted dual-beam phase-referencing technique and successfully detected the radio emission from M\,84 at 22 and 43\,GHz with VLBI Exploration of Radio Astrometry. They also found a jet-like structure extended to the north at 22\,GHz with a peak flux of 71\,mJy, while the peak flux at 43\,GHz was about 63\,mJy. 
                               
Among the nearby active galaxy nuclei, M\,84 is one of the rare sources with obvious two-side jets in the low-frequency VLBI images. Other similar sources such as Centaurus A \citep{Janssen} and NGC\,1052 \citep{Baczko}, own masses of one order of magnitude lower than M\,84. Previous VLBI studies on M\,84 focused on the basic structure and the emission of jet, without deeper studies and discussion on the kinematics and the morphology of the inner jet.
The source with a large viewing angle provides an opportunity to study the disk-jet connection near the event horizon. Such source will play an important role in understanding LLAGN with the improvement of the sensitivity and the angular resolution of VLBI.

In this paper, we present the multi-frequency and multi-epoch observations to probe the inner jet of M\,84. Our observational data and data reduction are described in section \ref{sec:obs}; The results and discussion about the proper motion, the viewing angle and the ridge structure are presented in section \ref{sec:results}. 
These results are summarized in section \ref{sec:summary}. 
We adopt an angular distance scale of 89\,pc/arcsec. For the supermassive black hole mass of $~8.5\times10^8$$M_\odot$ \citep{Walsh}, M\,84 has a Schwarzschild radius of $\sim$ 0.9\,$\mu$as. The apparent angular size of the black hole shadow of M\,84 assuming zero spin is 4.7\,$\mu$as \citep{Roelofs} (or even larger to 10\,$\mu$as under the mass of $~1.8\times10^9$$M_\odot$). The event horizon in M\,84 could be approached by future sub-mm VLBI technique \citep{Raymond, Jiang}.

\begin{figure*}[h]
	\centering
	\includegraphics[scale=0.7]{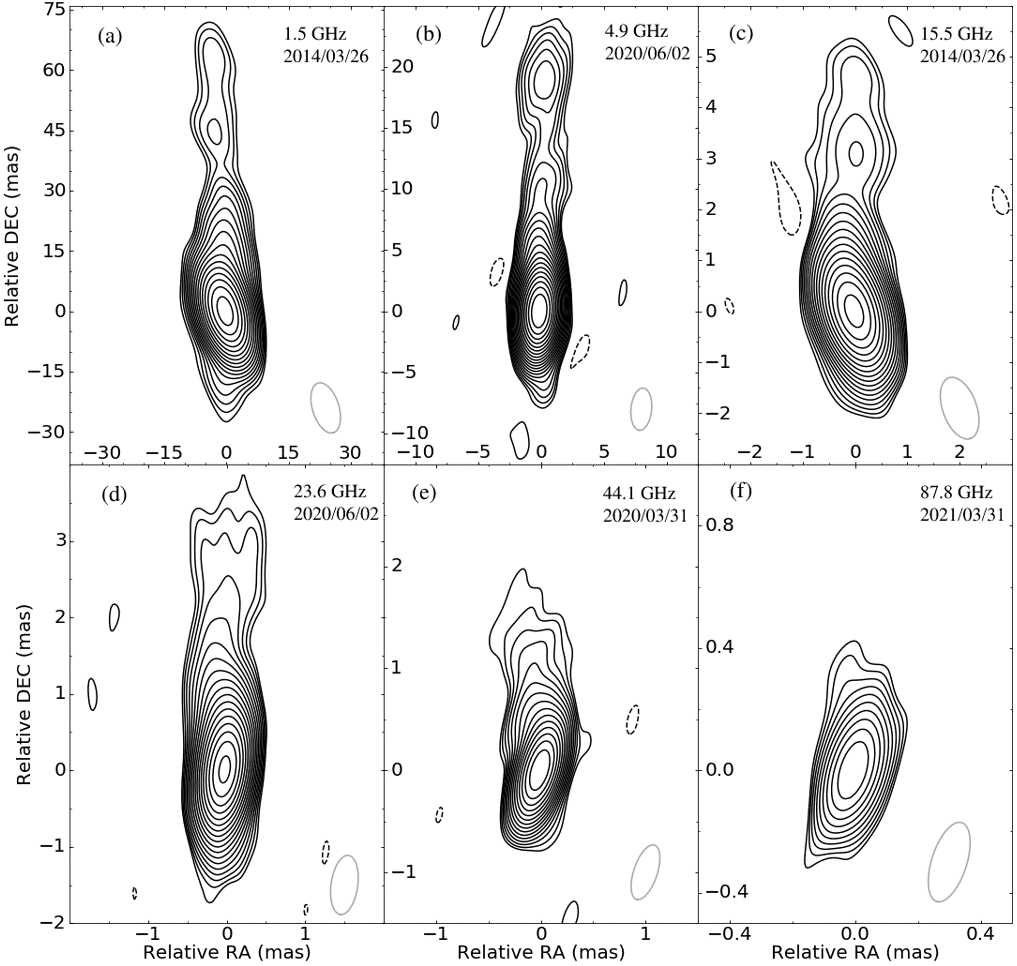}
	\caption{VLBI images of M\,84 with natural weighting from 1.5 to 88\,GHz. The parameters of images are shown in Table \ref{tab:image}. The contours are plotted at increasing powers of $\sqrt{2}$ from 3$\sigma$. The beam is plotted at the right lower corner. 
 \label{fig:image}  }
\end{figure*}	

\section{Observations and data reduction } \label{sec:obs}

We observed M\,84 with VLBI arrays including the VLBA at multi-frequency in three epochs and the East Asian VLBI Network (EAVN) at 22\,GHz in one epoch. Additionally, data in two epochs at 1.5, 5 and 15\,GHz from public VLBA archive system were re-analysed. The data we observed and re-analysed in this paper are listed in Table \ref{tab:image}.

\textbf{VLBA data}: The frequencies in VLBA epochs we observed cover 5, 22, 44 and 88\,GHz. All ten antennas participated in the observation of bx014 (the VLBA legacy experiment code and the same below), while bj094a was without the antenna FD and bj094b was without the antenna SC and HN. The data were recorded at recording rates of 2, 2 and 4\,Gbps. The total bandwidths were 512, 512, 1024\,MHz splitting into 8, 4, 8 intermediate frequency (IF) bands in bj094a, bx014 and bj094b, respectively. Bright sources M\,87 and 3C\,273 were used as the calibrators in all these epochs. 

We also analysed two epoch data in 2014 with an interval of 43\,days from NRAO science data archive at 1.5, 5 and 15\,GHz. These data were used for a complement of the spectrum analysis and the flux ratios of jet to counterjet.

\textbf{EAVN data}: We observed M\,84 with EAVN on 20 December 2019. There were nine antennas participating in the observation. It had an angular resolution of $\sim0.55$\,mas and an image sensitivity of $\sim75 \mu$\,Jy \citep{Cui}. The data were recorded at a recording rate of 1\,Gbps in left-hand circular polarization with 8\,IFs and each IF had a bandwidth of 32\,MHz.

The VLBA data were correlated by the VLBA correlator in Socorro, New Mexico. The EAVN data were correlated by the Daejeon correlator at the Korea-Japan Correlation Center \citep{Lee}. 
The amplitude and phase were calibrated using NRAO Astronomical Image Processing System \citep{Greisen}. The phase at 88\,GHz was calibrated with source-frequency phase-referencing and its image was reported in \citet{Jiang}.    
The data exported from AIPS were loaded into DIFMAP software package \citep{Shepherd}. The CLEAN and self-calibration procedures were used to obtain the final images. We present the self-calibrated and natural weighting images from 1.5 to 88\,GHz in Figure \ref{fig:image}.
The self-calibrated data were then fitted with circular Gaussian components through MODELFIT function to extract four parameters of each component (flux, distance from the core, position angle and component size).

The flux of core component in a19xw01a has been multiplied by a factor of 1.3 to correct the flux losses caused by the EAVN backend system and correlation \citep{Lee}. The self-calibrated fluxes in bj094a, bj094b and bx014 are scaled up in a factor of 1.15 because all VLBA data obtained after 2019 April 15 have suffered from a systematic decrease on all baselines \citep{Lister}. For the flux at 88\,GHz, we refer to the nearby calibrator source M\,87, whose core fluxes at 86\,GHz were reported in \citet{Kim}. We adopt its average flux in the stable period according to the light curve because no flare has been reported during our observation. A scaling factor of 2.17 is used and a corrected flux of $0.083 \pm 0.016$\,Jy is obtained for M\,84 at 88\,GHz.

In this paper, the errors of fluxes are estimated from the brightness temperatures \citep{Jorstad}: $\sigma_{\rm S} \approx 0.09 T_{\rm b,obs}^{-0.1}$ and the errors of positions are approximated as: $\sigma_{\rm X,Y} \approx 1.3 \times 10^4 T_{\rm b,obs}^{-0.6}$ plus a minimum error of one fifth angular resolution.

\section{Results and discussion} \label{sec:results}

\subsection{VLBI images} \label{subsec:images}

The VLBI images from 1.5 to 88\,GHz are shown in Figure \ref{fig:image}. The noise levels in new images are significantly improved comparing with previous studies because of the improvement of sensitivities. These images show richer structures such that we can identify the corresponding components in three epochs at 22\,GHz.
All images show a core-jet structure. The jet at 1.5\,GHz extends to about 60\,mas, or $\sim$5.3\,pc projected in the celestial plane, which shortens at higher frequencies due to the steep spectrum of the synchrotron radiation. These images, especially at frequencies less than 22\,GHz, show an elongated and collimated jet whose overall direction is almost due north while the counter-jet is southward. The jet directions at each frequency are consistent, including the image at 88\,GHz, in which the extended structure is down to a distance of $\sim 400\,R_s$ along the jet from the core.  
Such a morphology coincides with the outer radio jet observed by VLA and ALMA \citep{Meyer}. 
The jet is locally oscillating around the position angle of $0^{\circ}$, which is more obvious at higher angular resolution. Details on ridge structure are discussed in section \ref{subsec:helical}. 

\begin{figure}[h]
	\centering
	\includegraphics[scale=0.38]{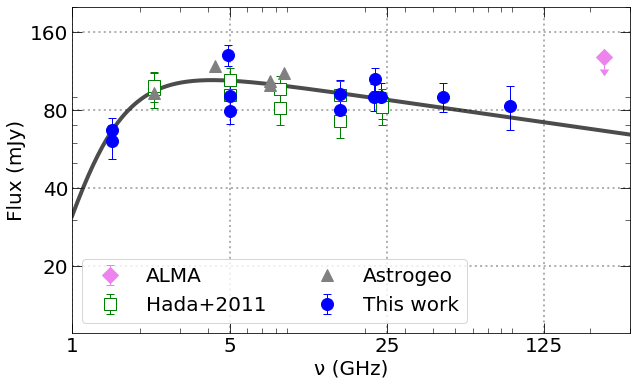}
	\caption{Spectrum of the M\,84 core. The blue circulars are the core fluxes of this work and the green squares are the core fluxes of Hada+2011 \citep{Hada}. The violet diamond is the data from ALMA observation as a reference in the submillimeter wavelength. The gray triangle is the compact flux from AstroGeo database. The black solid line is the fitted curve.
 \label{fig:spectrum}  }
\end{figure}	

The spectrum of core components from 1.5 to 88\,GHz is overall inverted-to-flat as shown in Figure \ref{fig:spectrum}. Some extra data points at 2.3, 5, 8.4 and 15.4\,GHz are quoted from \citet{Hada} denoted by Hada+2011, as well as the data from the Astrogeo VLBI FITS image database\footnote[1]{\href{http://astrogeo.org/vlbi_images}{http://astrogeo.org} since 2014 denoted by Astrogeo} for verification and complement of the spectral analysis. The data from ALMA at 230\,GHz \citep{Boizelle}, whose relatively higher flux is due to the low angular resolution and the blended emissions from the accretion disk \citep{Raiteri}, thus is referenced as an upper limit and not used for spectrum fitting. We fit the spectrum of the core following the flux-frequency equation described in \citep{Kosogorov} from 1.5 to 88\,GHz. The spectral index in the low frequency is set as +2.5 due to the synchrotron radiation self-absorption. We obtain an inverted-to-flat curve and a spectral index of $\alpha_{\rm c} = -0.12 \pm 0.02$ for the flat spectrum, which is consistent with the result reported in \citet{Nakahara}. The estimated turnover frequency is $\nu_{\rm m} = 4.2\pm0.2$\,GHz and the corresponding flux is $F_{\rm m} = 0.105\pm 0.010$\,Jy. We obtain a magnetic field strength of the order of 1$\sim$10\,mG and an electron density of $\sim 10^5\,\rm cm^{-3}$ following the equations  in \citet{Hirotani} with an estimated angular diameter of the core $0.28\pm0.03$\,mas, which is derived from the fitted relation between the core size and the frequency. 

The components in the counterjet are found in most of images. Even in the image at 88\,GHz with large RMS, the core appears to be slightly extended to the south.

\begin{deluxetable*}{ccccc}
\tablenum{2}
\tablecaption{Gaussian component parameters at 22\,GHz \label{tab:component}}
\tablewidth{0pt}
\tablehead{
\colhead{Component} & \colhead{Interval (day)} & \colhead{Distance (mas)}  & \colhead{P.A. (deg)}  &\colhead{FWHM (mas)}
}
\decimalcolnumbers
\startdata
\multirow{3}{*}{Na}  & 0 	& 0.58$\pm$0.16  & 4.4  & 0.03 \\
							 & 181& 0.76$\pm$0.17 &  7.9 		& 0.03 \\
 						    &346 & 0.96$\pm$0.31 &  2.1  & 0.17 \\
\hline
\multirow{3}{*}{Nb} &0 & 1.25$\pm$0.46 & 2.7 & 0.30 \\
							 & 181 & 1.58$\pm$0.46 & 6.1 & 0.27 \\
 							& 346 & 2.13$\pm$0.64 & $-$5.3 &0.34 \\
\hline    
\multirow{3}{*}{Nc} & 0 & 2.22$\pm$0.41 & 5.6  & 0.16 \\
							 & 181 & 2.80$\pm$0.53 & $-$2.5  & 0.26 \\
 							& 346& 3.20$\pm$1.43 & 1.9 & 0.86 \\             
\enddata
\tablecomments{Parameters of Gaussian components in three epochs at 22\,GHz. The distance from the core, position angle and FWHM of circular Gaussian are listed by number of columns.}
\end{deluxetable*} 
\begin{figure}[h]
	\centering
	\includegraphics[scale=0.37]{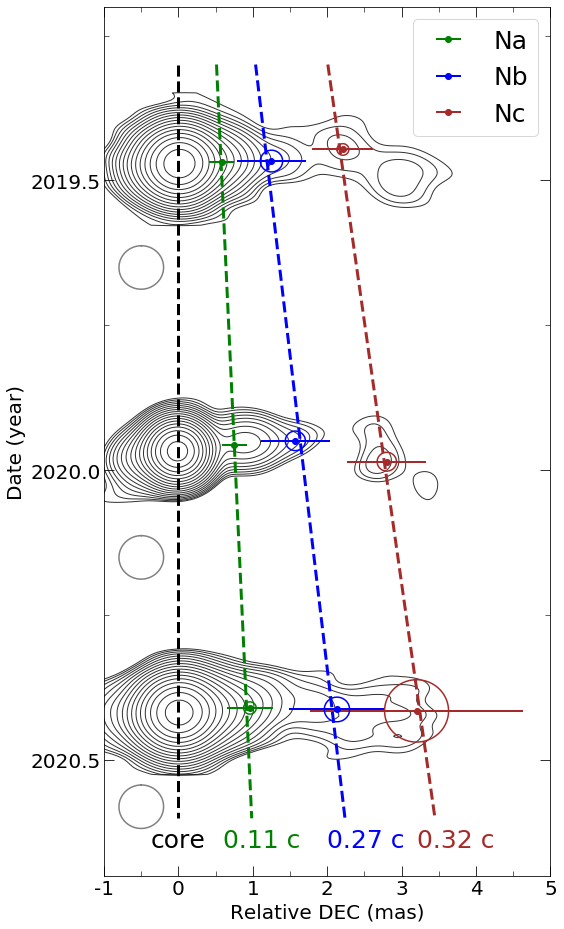}
	\caption{Time evolution of M\,84 at 22\,GHz. Green, blue and red (or circulars) represent the components of Na, Nb and Nc, respectively. The dotted lines are used to fit their speeds. The contours are plotted at increasing powers of $\sqrt{2}$ from 0.3\,mJy. The circular beam is plotted at the left lower corner. \label{fig:motion} }
\end{figure}

\begin{figure}[h]
\plotone{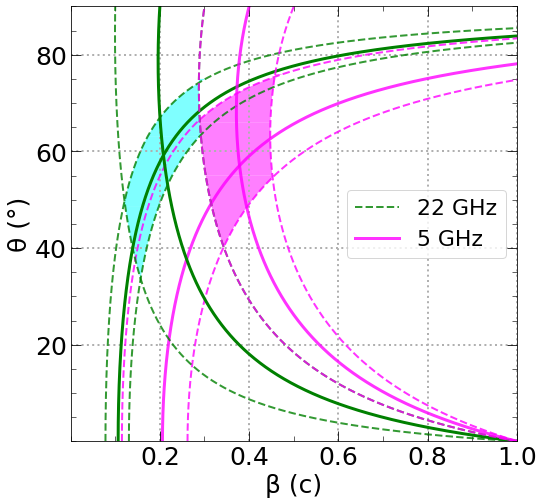}
\caption{Constraints on the values of the intrinsic jet speed and the angle to the line of sight. The green lines use the flux ratios of $1.9\pm0.3$ and apparent speeds of $0.2\pm0.1$\,c, respectively. The magenta lines use the ratios of $3.5\pm1.5$, and an apparent speed of $0.4\pm0.1$\,c. The cyan and magenta regions are the viewing angle results. The spectral index of jet $\alpha_{\rm j}$ in Equation (1) is assumed to be $-$1.    
\label{fig:angle}}
\end{figure}

\subsection{Jet proper motion and viewing angle} \label{subsec:angle}

The jet proper motion was monitored at 22\,GHz in three epochs throughout one year. The proper motion between bh186d and bh186e at 15\,GHz is checked. The components in these two epochs with an interval of 43 days are one to one correspondence with slight position displacements, which means the apparent speed is not too large to cause a dislocation of the components. Therefore, the adjacent components at 22\,GHz in three epochs separated every six months could be recognized as the same components. The displacements of components are obvious in both the model-fitting and the bright features in the images as shown in Figure \ref{fig:motion}.

Three components named Na, Nb and Nc are identified in the jet, whose position parameters are listed in Table \ref{tab:component} and their increasing separations from the core over time are plotted in Figure \ref{fig:motion}.
We also examine the flux profile along the ridge line, where we can find the corresponding components obtained from model-fitting. The final fitting result of proper motions still has non-negligible uncertainties and a risk in calculating the apparent speeds. The uncertainties are mainly from the insufficient cadence of epochs and the model-fitting.
The fitted proper motions are about 0.36, 0.93, and 1.11\,mas/yr, equivalent to apparent speeds of 0.11\,c, 0.27\,c, and 0.32\,c, respectively. These speeds are much less than that in the outer jet, $>$ 3\,mas/yr \citep{Meyer}.

A counterjet towards south is obviously seen at frequencies below 22\,GHz, whose direction is consistent with the kpc-scale counterjet. For the two-side jet, the flux ratio of jet to counterjet is used to calculate the viewing angle of the inner jet:
\begin{equation}
R_{\rm flux} = \left({\frac{1+\beta \cos\theta}{1-\beta \cos\theta}}\right)^{2-\alpha_{\rm j}},
\end{equation}
where $\alpha_{\rm j}$ is the spectral index of the jet; $\beta$ is the intrinsic speed in unit of c and $\theta$ is the viewing angle. 
We restore the images with a circular beam and compute the flux ratio between slices which are equidistant from the core. The cutoff flux of counterjet larger than $15 \sigma$ along the ridge line is selected in bx014. We calculate the ratios in regions of $4.0\sim5.4$\,mas at 5\,GHz and $0.9\sim1.2$\,mas at 22\,GHz, where the ratios are $2.1\sim5.0$ and $1.6\sim 2.2$, respectively. The former region is farther than Nc, whose increasing flux ratio with distance from the core indicates that the jet is accelerating, thus has a larger apparent speed estimated as $0.4\pm0.1$\,c. The latter one is between components Na and Nb, so its apparent speed is estimated as $0.2\pm0.1$\,c.
Combining Equation (1) with the equation of apparent speed:
\begin{equation}
\beta_{\rm app} = \frac{\beta \sin\theta}{1-\beta \cos\theta} ,
\end{equation}
we could constrain the viewing angle under the assumption of intrinsic symmetry between the jet and the counterjet.
The spectral index of jet $\alpha_{\rm j}$ in the Equation\,(1) is presumed as the normal value of $-$1. 
As shown in Figure \ref{fig:angle}, the viewing angle is ${59^{+16}_{-26}}^{\circ}$ calculated at 22\,GHz and ${57^{+19}_{-17}}^{\circ}$ at 5\,GHz. These two results are similar, and are both smaller than the result from outer jet (${74^{+9}_{-18}}^{\circ}$) \citep{Meyer}. The overlapping angle from two results is ${58^{+17}_{-18}}^{\circ}$.

The intrinsic speeds of the components in Table \ref{tab:component} after applying the viewing angle of $58^{\circ}$ to Equation (2) are 0.12\,c, 0.27\,c, and 0.32\,c, at an average de-projected distance of 980\,$R_{\rm s}$, 2100\,$R_{\rm s}$, and 3700\,$R_{\rm s}$, respectively. 
The distribution of intrinsic speeds versus distance from the core in unit of  $R_{\rm s}$ is at a same level as that in NGC\,1052 \citep{Baczko}.

\subsection{Jet Oscillation} \label{subsec:helical}
\begin{figure*}[htb]
	\centering
	\includegraphics[scale=0.5]{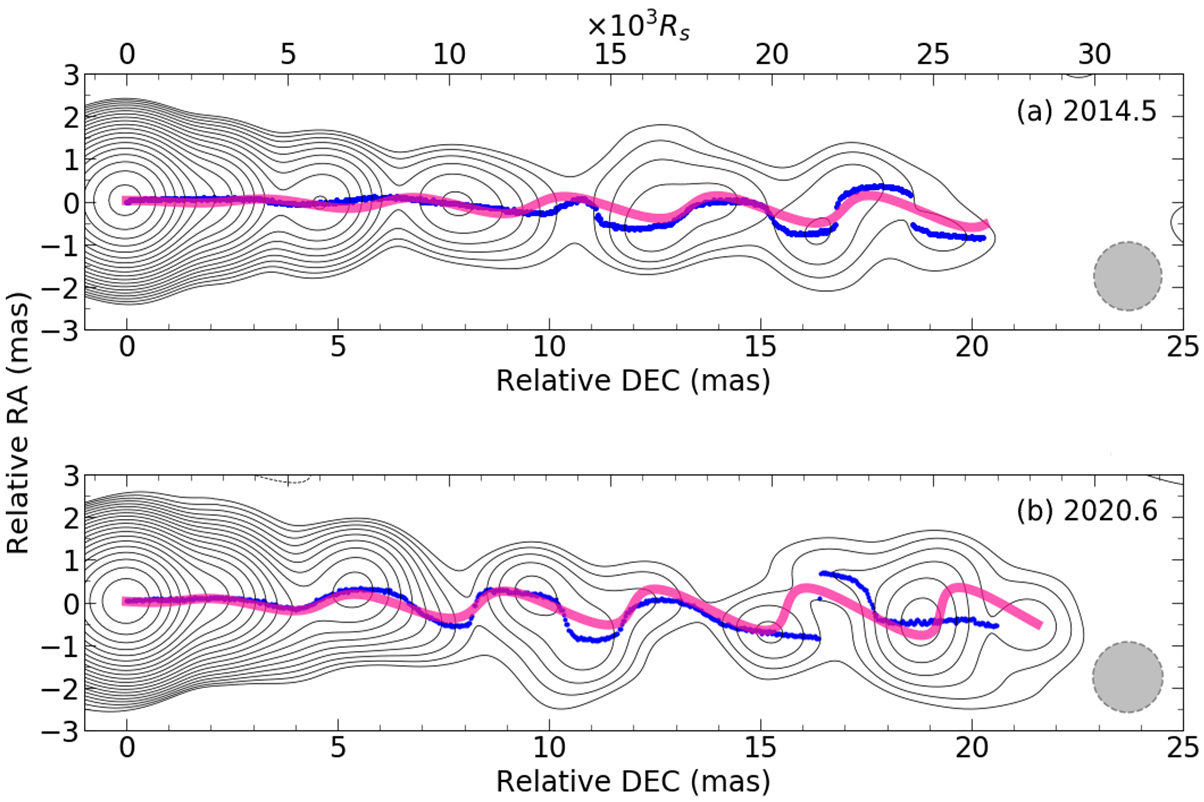}
	\caption{Illustration of the model-fitting between the ridge structures and the wave models of the Kelvin-Helmholtz instability along the jet in bh186e (a) and bx014 (b). The blue dotted lines are the ridge lines and the pink solid lines are the corresponding models. The images are restored under natural weighting scheme with a circular beam as shown in the right lower corner. The upper axis is the de-projected count of Schwarzschild radius. \label{fig:helical}}
\end{figure*}

It is noticed the jet morphology in M\,84 is oscillatory indicated from the changing orientation angle of the ridge line. We mainly focus on the images at 5\,GHz, where the jet structure is reconstructed with a high sensitivity and thus a high dynamical range. The ridge lines of wave-like morphology in bx014 and bh186e restored with a circular beam under natural weighting are shown in Figure \ref{fig:helical}. In order to exclude the possibility that an oscillatory morphology might be caused by the sidelobes when we carry out the CLEAN and self-calibration procedures, we check the ridge line in the image under uniform weighting scheme, and find that the directions of ridge line is consistent with the one under natural weighting scheme in Figure \ref{fig:helical}. Moreover, such morphology is found in both bx014 and bh186e with a similar oscillatory wavelength. Thus, the oscillation in M\,84 is considered as a true structure.  

An oscillatory jet is generally considered to be caused by the precession of central engine or by Kelvin-Helmholtz (KH) instability. For instance, a binary black hole system can generate the precession of jet due to the orbital motion of the primary black hole \citep{Villata}, but M\,84 is an elliptical galaxy and has no obvious clues about the existence of binary black hole. KH instability is another general interpretation of the jet oscillation in many sources, such as 3C\,273 \citep{Lobanovb}, S5\,0836+710 \citep{Vega}, M\,87 \citep{Hardee}. 

Following the model in \citet{Lobanovb}, the jet oscillation can be modelled by a combination of several helical waves. In our model, a single apparent pattern with a wavelength $\lambda_{\rm obs} \approx3.4$\,mas is used to characterize the oscillation of ridge line. Figure \ref{fig:helical} shows the fitting results between the jet ridge and a model of single sinusoidal mode after $58^{\circ}$ projection in two epochs. Some discrepancies between the model and ridge lines can be interpreted as the enlarged errors in the further away from the center of image or the growing amplitudes of some other potential wave modes of KH instability. 
It is rare to find a single wave pattern of KH instability to characterize the jet oscillation. A superpositon of several oscillatory patterns of KH instability is common in other sources \citep{Vega}. An explanation is that one pattern of KH instability is dominated while other patterns are still growing and do not fully appear, especially considering that the observed region ($10^3\sim10^4 R_{\rm s}$) in M\,84 is the innermost jet.
We find that the improvement of angular resolution will lead to an increased amplitude when comparing the images under different beam sizes, while the phase along the propagation direction does not change. This is reasonable that the flux distribution would be smoothed at a low resolution. The amplitude will reach an upper limit when the transverse jet is completely resolved. We check the image restored under uniform weighting, whose ridge line is gradually tend to more coincide with that in Figure \ref{fig:helical} (b) along the jet. The jet in the downstream region could be considered to be completely resolved.

The model also works in the image at 22\,GHz in bx014 with only an initial phase difference of $\sim 40^{\circ}$ from the simultaneous image at 5\,GHz (Figure \ref{fig:helical}b). If this difference is caused by the core shift, its approximate value between 5 and 22\,GHz would be 0.38\,mas. The magnetic field strength estimated from this core-shift value following \citet{Kosogorov} is also at the same level of serveral mG. 

The Mach number in the jet $M_{\rm j}$ and in the ambient medium $M_{\rm x}$, as well as the the particle density ratio of the jet to the ambient medium $\eta = M^{\rm 2}_{\rm j}/M^{\rm 2}_{\rm x}$ can be calculated with accurate values of the characteristic wavelength, the jet radius, the viewing angle, the jet apparent speed, and the apparent pattern (or wave) speed \citep{Vega}. For the surface and helical mode, the characteristic wavelength is $1.5\lambda_{\rm obs}$. The only parameter that we could not determine is the apparent pattern speed. The phase difference between two images at 5\,GHz (Figure \ref{fig:helical}) is $\sim 130^{\circ}$. It indicates that the wave has propagated at least $130^{\circ}$ over 6 years. If the wave has propagated $n$ periods, the apparent pattern speed will be about $0.06+0.16n$\,c. Dedicated dense observations are required to measure the apparent pattern speed in the future.                                

\section{Summary} \label{sec:summary}

In this paper, we report detailed analysis on the inner jet/counterjet of M84 through multi-frequency and multi-epoch observations. 
We present the VLBI images from 1.5 to 88\,GHz, whose morphologies show core-jet structures.
We fit the fluxes in core components and obtain an inverted-to-flat spectrum from 1.5 to 88\,GHz, whose magnetic field strength and electron number density are estimated based on the turnover frequency.

The apparent speeds for three components near the core are about 0.11\,c, 0.27\,c, and 0.32\,c. These different speeds and the distribution of flux ratios of jet to counterjet indicate that a significant acceleration is occurring in the inner region. This result is consistent with the general acceleration scale in other AGNs. Considering the relatively large viewing angle of ${58^{+17}_{-18}}^{\circ}$, the intrinsic speeds are moderately relativistic. 

We use a single wave mode of the Kelvin-Helmholtz instability to explain the sinusoidal-like jet morphology at 5\,GHz. The model with an apparent wavelength of about 3.4\,mas can well fit the ridge line in two images. This wave-like ridge line and Kelvin-Helmholtz instability model in M\,84 should be examined with further dedicated observations.

\begin{acknowledgments}
We appreciate the important and motivating advice from the anonymous referee, which helped to improve the manuscript significantly. This work was supported in part by the National Natural Science Foundation of China (grant Nos. 12173074, 11803071, 11933007), the Key Research Program of Frontier Sciences, CAS (grant Nos. QYZDJ-SSW-SLH057, ZDBS-LY-SLH011). We thank the observation facilities VLBA, EAVN (including TMRT, NSRT, KVN and VERA), and all people working for observation and correlation. TMRT is operated by Shanghai Astronomical Observatory. NSRT is operated by Xinjiang Astronomical Observatory. KVN is a facility operated by the Korea Astronomy and Space Science Institute. VERA is a facility operated at National Astronomical Observatory of Japan. We acknowledge the MOJAVE database for querying the data, which is maintained by the MOJAVE team. This study makes use of 43\,GHz VLBA data from the VLBA-BU Blazar Monitoring Program, funded by NASA through the Fermi Guest Investigator Program. We wolud like to thanks the VLBA data archive maintained by NRAO for the essential data. VLBA and NRAO are the facilities of the National Science Foundation operated under cooperative agreement by Associated Universities, Inc. 
\end{acknowledgments}

%

\vspace{5mm}





\bibliography{sample631}{}

\begin{thebibliography}{}

%


\bibitem[Baczko et al.(2019)]{Baczko} Baczko, A.-K., Schulz, R., Kadler, M., et al.\ 2019, \aap, 623, A27. doi:10.1051/0004-6361/201833828
\bibitem[Boizelle et al.(2017)]{Boizelle} Boizelle, B.~D., Barth, A.~J., Darling, J., et al.\ 2017, \apj, 845, 170. doi:10.3847/1538-4357/aa8266
\bibitem[Cui et al.(2021)]{Cui} Cui, Y.-Z., Hada, K., Kino, M., et al.\ 2021, Research in Astronomy and Astrophysics, 21, 205. doi:10.1088/1674-4527/21/8/205
\bibitem[Greisen(2003)]{Greisen} Greisen, E.~W.\ 2003, Information Handling in Astronomy - Historical Vistas, 285, 109. doi:10.1007/0-306-48080-8\_7
\bibitem[Hada et al.(2011)]{Hada} Hada, K., Doi, A., Kino, M., et al.\ 2011, \nat, 477, 185. doi:10.1038/nature10387
\bibitem[Hardee \& Eilek(2011)]{Hardee} Hardee, P.~E. \& Eilek, J.~A.\ 2011, \apj, 735, 61. doi:10.1088/0004-637X/735/1/61
\bibitem[Hirotani(2005)]{Hirotani} Hirotani, K.\ 2005, \apj, 619, 73. doi:10.1086/426497
\bibitem[Janssen et al.(2021)]{Janssen} Janssen, M., Falcke, H., Kadler, M., et al.\ 2021, Nature Astronomy, 5, 1017. doi:10.1038/s41550-021-01417-w
\bibitem[Jiang et al.(2021)]{Jiang} Jiang, W., Shen, Z., Mart{\'\i}-Vidal, I., et al.\ 2021, \apjl, 922, L16. doi:10.3847/2041-8213/ac375c
\bibitem[Jorstad et al.(2017)]{Jorstad} Jorstad, S.~G., Marscher, A.~P., Morozova, D.~A., et al.\ 2017, \apj, 846, 98. doi:10.3847/1538-4357/aa8407
\bibitem[Kim et al.(2018)]{Kim} Kim, J.-Y., Krichbaum, T.~P., Lu, R.-S., et al.\ 2018, \aap, 616, A188. doi:10.1051/0004-6361/201832921
\bibitem[Kosogorov et al.(2022)]{Kosogorov} Kosogorov, N.~A., Kovalev, Y.~Y., Perucho, M., et al.\ 2022, \mnras, 510, 1480. doi:10.1093/mnras/stab3579
\bibitem[Lee et al.(2015)]{Lee} Lee, S.-S., Oh, C.~S., Roh, D.-G., et al.\ 2015, Journal of Korean Astronomical Society, 48, 125. doi:10.5303/JKAS.2015.48.2.125
\bibitem[Lister et al.(2021)]{Lister} Lister, M.~L., Homan, D.~C., Kellermann, K.~I., et al.\ 2021, \apj, 923, 30. doi:10.3847/1538-4357/ac230f
\bibitem[Lobanov \& Zensus(2001)]{Lobanovb} Lobanov, A.~P. \& Zensus, J.~A.\ 2001, Science, 294, 128. doi:10.1126/science.1063239
\bibitem[Ly et al.(2004)]{Ly} Ly, C., Walker, R.~C., \& Wrobel, J.~M.\ 2004, \aj, 127, 119. doi:10.1086/379855
\bibitem[Meyer et al.(2018)]{Meyer} Meyer, E.~T., Petropoulou, M., Georganopoulos, M., et al.\ 2018, \apj, 860, 9. doi:10.3847/1538-4357/aabf39
\bibitem[Nagar et al.(2002)]{Nagar} Nagar, N.~M., Falcke, H., Wilson, A.~S., et al.\ 2002, \aap, 392, 53. doi:10.1051/0004-6361:20020874
\bibitem[Nakahara(2014)]{Nakahara} Nakahara, S.\ 2014, Proceedings of the 12th European VLBI Network Symposium and Users Meeting (EVN 2014). 7-10 October 2014. Cagliari, 100
\bibitem[O'Sullivan \& Gabuzda(2009)]{O'Sullivan} O'Sullivan, S.~P. \& Gabuzda, D.~C.\ 2009, \mnras, 400, 26. doi:10.1111/j.1365-2966.2009.15428.x
\bibitem[Raiteri et al.(2014)]{Raiteri} Raiteri, C.~M., Villata, M., Carnerero, M.~I., et al.\ 2014, \mnras, 442, 629. doi:10.1093/mnras/stu886
\bibitem[Raymond et al.(2021)]{Raymond} Raymond, A.~W., Palumbo, D., Paine, S.~N., et al.\ 2021, \apjs, 253, 5. doi:10.3847/1538-3881/abc3c3
\bibitem[Roelofs et al.(2019)]{Roelofs} Roelofs, F., Falcke, H., Brinkerink, C., et al.\ 2019, \aap, 625, A124. doi:10.1051/0004-6361/201732423
\bibitem[Shepherd(1997)]{Shepherd} Shepherd, M.~C.\ 1997, Astronomical Data Analysis Software and Systems VI, 125, 77
\bibitem[Vega-Garc{\'\i}a et al.(2020)]{Vega} Vega-Garc{\'\i}a, L., Lobanov, A.~P., Perucho, M., et al.\ 2020, \aap, 641, A40. doi:10.1051/0004-6361/201935168
\bibitem[Villata \& Raiteri(1999)]{Villata} Villata, M. \& Raiteri, C.~M.\ 1999, \aap, 347, 30
\bibitem[Walsh et al.(2010)]{Walsh} Walsh, J.~L., Barth, A.~J., \& Sarzi, M.\ 2010, \apj, 721, 762. doi:10.1088/0004-637X/721/1/762







\end{thebibliography}
\bibliographystyle{aasjournal}



\end{document}